\documentclass[final,5p,times,twocolumn,fleqn]{elsarticle}

\usepackage{chemformula}
\let\ce\ch

\usepackage{subcaption}

\usepackage{threeparttable}
\usepackage{etoolbox}
\appto\TPTnoteSettings{\footnotesize}

\usepackage{tabularx}
\usepackage{booktabs}

\usepackage{multirow}
\usepackage{makecell}

\usepackage[colorlinks=true]{hyperref}
\usepackage[capitalise,nameinlink,noabbrev]{cleveref}
\usepackage{natbib}

\journal{Nuclear Instruments and Methods in Physics Research - section A}

\begin{document}

\begin{frontmatter}

\title{Energy calibration of GTM on ground}

\author[inst1]{Chien-You Huang\corref{cor1}}
\ead{ak119ka@gmail.com}
\author[inst1,inst2,inst3]{Hsiang-Kuang Chang\corref{cor1}}
\ead{hkchang@mx.nthu.edu.tw}
\cortext[cor1]{Corresponding author}

\author[inst4]{Chih-Hsun Lin}
\author[inst5]{Che-Chih Tsao}
\author[inst6]{Chin-Ping Hu}
\author[inst1]{Hao-Min Chang}
\author[inst5]{Yan-Fu Chen}
\author[inst5]{An-Hsuan Feng}
\author[inst6]{Yi-Wen Huang}
\author[inst1]{Tzu-Hsuan Lin}
\author[inst7]{Yi-Ning Tsao}
\author[inst2]{Chih-En Wu}
\author[inst5]{Chun-Wei Wu}

\affiliation[inst1]{organization={Department of Physics},
                    addressline={National Tsing Hua University},
                    city={Hsinchu},
                    postcode={300044},
                    country={Taiwan}}
\affiliation[inst2]{organization={Institute of Astronomy},
                    addressline={National Tsing Hua University},
                    city={Hsinchu},
                    postcode={300044},
                    country={Taiwan}}
\affiliation[inst3]{organization={Institute of Space Engineering},
                    addressline={National Tsing Hua University},
                    city={Hsinchu},
                    postcode={300044},
                    country={Taiwan}}
\affiliation[inst4]{organization={Institute of Physics},
                    addressline={Academia Sinica},
                    city={Taipei},
                    postcode={115201},
                    country={Taiwan}}
\affiliation[inst5]{organization={Department of Power Mechanical Engineering},
                    addressline={National Tsing Hua University},
                    city={Hsinchu},
                    postcode={300044},
                    country={Taiwan}}
\affiliation[inst6]{organization={Department of Physics},
                    addressline={National Changhua University of Education},
                    city={Changhua},
                    postcode={500207},
                    country={Taiwan}}
\affiliation[inst7]{organization={Arete Honors Program},
                    addressline={National Yang Ming Chiao Tung University},
                    city={Hsinchu},
                    postcode={300093},
                    country={Taiwan}}

\begin{abstract}
The Gamma-ray Transients Monitor (GTM) on board the Formosat-8B (FS-8B) satellite is designed to detect and localize Gamma-Ray Bursts (GRBs). By utilizing $2+2$ CITIROC chips to manipulate $4+4$ detectors, which are composed of \ce{GAGG(Ce)} scintillators coupled with Silicon Photomultipliers (SiPMs) and oriented in various directions to achieve all-sky coverage, the GRB saturation fluences of GTM in the 50 keV to 1 MeV range for Short GRBs (SGRBs) and Long GRBs (LGRBs) were estimated to be about $3.1 \times 10^{-4}$ and $5.0 \times 10^{-3}\ {\rm erg/cm^2}$, respectively, based on simulations. To precisely interpret the GTM readout signal in terms of energy, several measurements for isotope and gain calibration were conducted. Despite encountering issues with crosstalk and SiPM saturation effect in the data, the energy spectrum can still be recovered by appropriately discarding channel noise and mapping with the correct ADC-to-energy relation. This paper summarizes the energy resolution of GTM and the linear variations in the relationship between photon energy and readout signal. At 662 keV, the energy resolution is about $16\ \%$. Also, it demonstrates that greater gain is achieved by increasing voltage or decreasing temperature.
\end{abstract}


\end{frontmatter}

\section{Introduction}
\label{sec:1}

Since the first Gamma-Ray Burst (GRB) was discovered \citep{Klebesadel_1973}, numerous missions have observed thousands of GRBs over the past half-century \citep{Paciesas_1999,Tsvetkova_2017,Frontera_2008,Lien_2016,vonKienlin_2022}. The afterglows of GRBs were detected through the multi-wavelength observations \citep{Costa_1997,Van_1997,Frail_1997}. The precise location determined by the afterglow allowed for the identification of the host galaxy of GRB, from which the redshift can be inferred. Based on the bimodal distribution of the durations, GRBs were phenomenologically categorized into Short GRBs (SGRBs) and Long GRBs (LGRBs) \citep{Kouveliotou_1993}. In the mainstream theoretical framework \citep{Meszaros_2002}, the progenitor of LGRBs is attributed to the core collapse of a massive star \citep{Woosley_2006}, while SGRBs are believed to originate from the merger of binary compact stars \citep{Abbott_2017}. However, the duration classification is not always valid \citep{Gehrels_2006,Levesque_2010}. Other redshift-dependent quantities, such as the isotropic energy ($E_{iso}$), the intrinsic spectral peak energy ($E_{p,i}$), and the peak luminosity ($L_{p}$), have been used to classify GRB types and grasp their physical mechanisms more comprehensively \citep{Lu_2010,Zhang_2021}. The strong correlations among GRB quantities, such as the Amati relation \citep{Amati_2002}, Ghirlanda relation \citep{Ghirlanda_2004}, and Yonetoku relation \citep{Yonetoku_2004}, provided a means to probe the early universe \citep{Izzo_2015,Demianski_2017}, thanks to the wide-range redshift of GRBs. Consequently, more space telescopes monitoring GRBs are necessary to unravel the mysteries of GRBs and the fundamentals of the universe.

To help achieve the goal of expanding the sky coverage and enhancing the localization accuracy by synergizing with other telescopes, which can increase the probability of capturing afterglows for obtaining redshift information, the Gamma-ray Transients Monitor (GTM) is designed to monitor GRBs \citep{Chang_2022}. GTM on board Formosat-8B (FS-8B) satellite, manufactured by the Taiwan Space Agency (TASA), is scheduled to be launched into the Sun-Synchronous Orbit (SSO, with altitude = 561 km and inclination = $97.64\ ^\circ$) in 2026. As GTM is a secondary science payload, it can only operate while FS-8B is in the eclipse. Hence, the duty cycle is about $36\ \%$. According to the simulation with Medium-Energy Gamma-ray Astronomy library (MEGAlib) \citep{Zoglauer_2008}, the detection efficiency exceeds $50\ \%$ when the GRB fluences (in the energy range from 10 keV to 1 MeV) are greater than $6 \times 10^{-7}$ and $2 \times 10^{-6}\ {\rm erg/cm^2}$ for SGRBs and LGRBs, respectively. Referring to the distribution of Fermi/GBM 10-year GRB fluences and considering the duty cycle, GTM is anticipated to detect roughly 50 GRBs per year. Moreover, regarding localization capability, the uncertainty at the 3 sigma confidence level for LGRBs with fluences of $4 \times 10^{-6}$ and $4 \times 10^{-5}\ {\rm erg/cm^2}$ are about 30 and 3 degrees, respectively \citep{Huang_2024}.

To detect gamma-ray photons, GTM utilizes the Gadolinium Aluminium Gallium Garnet doped with Cerium [\ce{GAGG(Ce)}] scintillators and Silicon Photomultipliers (SiPMs) to convert the incident energy into electronic readout signals. The scintillation process occurs when an energetic photon or particle collides with a \ce{GAGG(Ce)} crystal, causing an inner-shell electron to absorb the energy and become excited into the conduction band, leaving behind a hole in its original position \citep{Rodnyi_1997,Bizarri_2010}. This primary electron-hole pair, in a non-equilibrium state, can induce various physical processes that contribute to the aggregation of many electron-hole pairs at the bottom of the conduction band and the top of the valence band. These pairs then fall into the luminescence center, created by the \ce{Ce} dopant, and emit a large amount of scintillating light to return to equilibrium. The light can be radiated in any direction, so Barium sulfate (\ce{BaSO4}), which acts as a mirror, effectively directs them into a SiPM. There are 14,336 independent microcells inside a single SiPM ($6\ {\rm mm} \times 6\ {\rm mm}$), called Single-Photon Avalanche Diodes (SPADs, each $50\ {\rm \mu m} \times 50\ {\rm \mu m}$). When a scintillating photon hits a SPAD, an avalanche process is fired to amplify this incident single photon \citep{Acerbi_2019}. With a quenching resistor in series with a SPAD, the avalanche phenomenon is terminated, allowing the SPAD to return to its initial state and wait for the next trigger. To comprehend the energy characteristics of GTM, calibration measurements are required.

This paper presents additional details about the instrument and energy calibration tests as follows. In \cref{sec:2}, we describe the electronics of the assembled GTM and estimate the saturation fluence of GRBs. In \cref{sec:3}, we explain the methods and settings used for the calibration measurements. In \cref{sec:4}, we outline the strategy for data reduction, which addresses and resolves the crosstalk issue. Lastly, we organize and discuss the results related to the energy characteristics of GTM in \cref{sec:5} and provide a summary in \cref{sec:6}.

\section{Instrument}
\label{sec:2}

\subsection{Module and Detector}
\label{sec:2.1}

\begin{figure}[t]
    \centering
    \includegraphics[width=.9\linewidth]{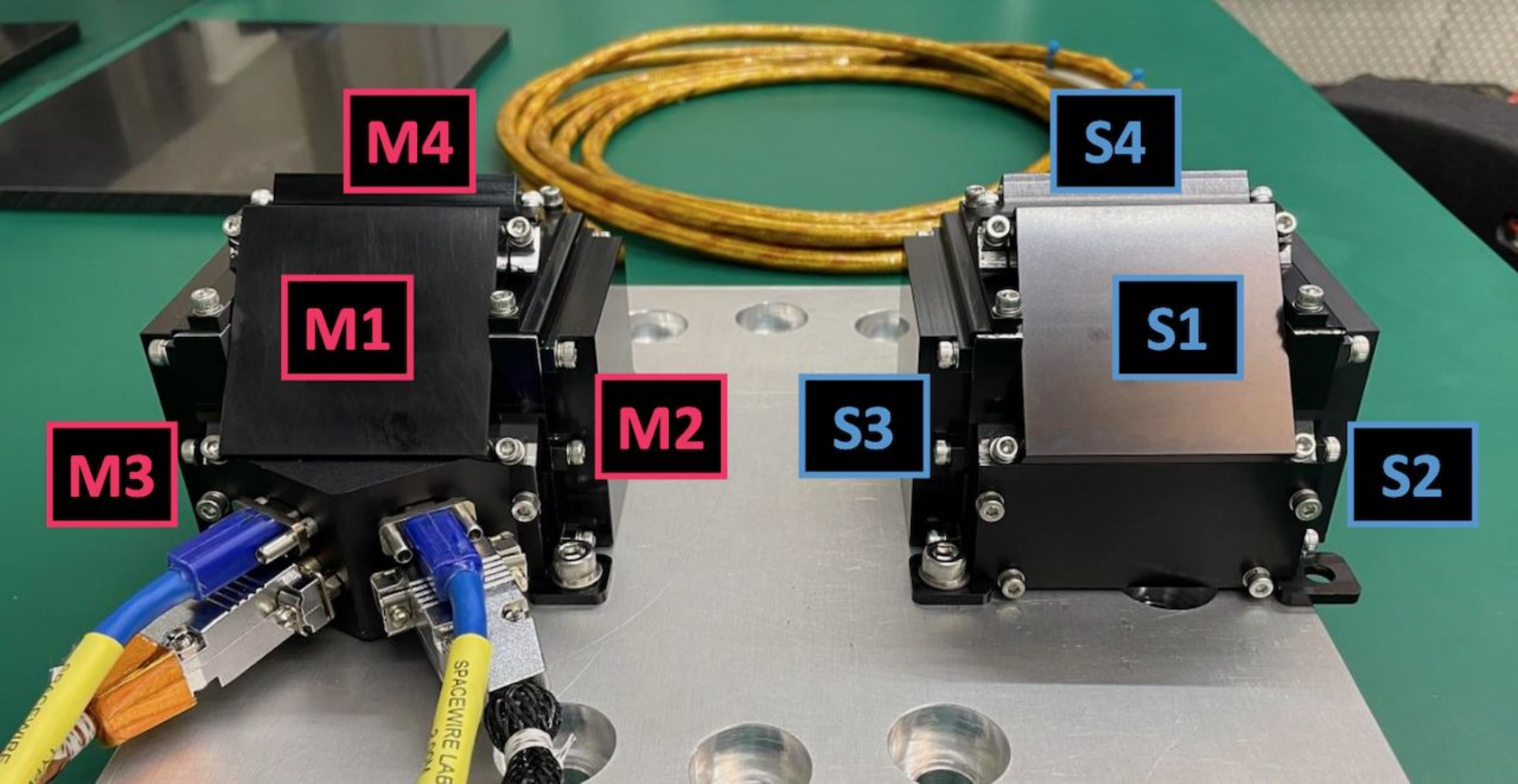}
    \caption{GTM/FM and its detector names and positions. The Master (M) and Slave (S) modules are labeled in red and blue colors, respectively. 4 detectors within each module are numbered.}
    \label{fig:1}
\end{figure}

\cref{fig:1} displays the master and slave modules of GTM's Flight Model (FM). Each module has 4 detectors pointed in different directions. To reduce the background noise generated by low-energy incident photons or particles, each detector is covered by an aluminum shell (1 mm thick at the top and 2 mm thick on 4 sides) that needs to be anodized to prevent the generation of static electricity. This causes GTM to appear black in color. Two modules will be placed on either side of FS-8B. Therefore, GTM can monitor GRBs across the entire sky by using the time variation of the count rate \citep{Goldstein_2020}. Additionally, the most probable location of GRB can also be fitted by using the ratio between the count rates of the 8 detectors \citep{Connaughton_2015}.

\begin{figure}[t]
    \centering
    \begin{subfigure}{.48\linewidth}
        \centering
        \includegraphics[width=.95\linewidth]{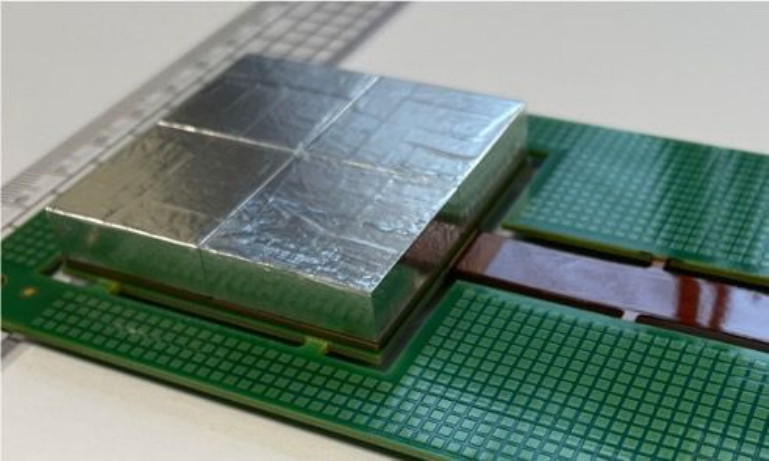}
        \caption{}
        \label{fig:2.1}
    \end{subfigure}
    \begin{subfigure}{.48\linewidth}
        \centering
        \includegraphics[width=.95\linewidth]{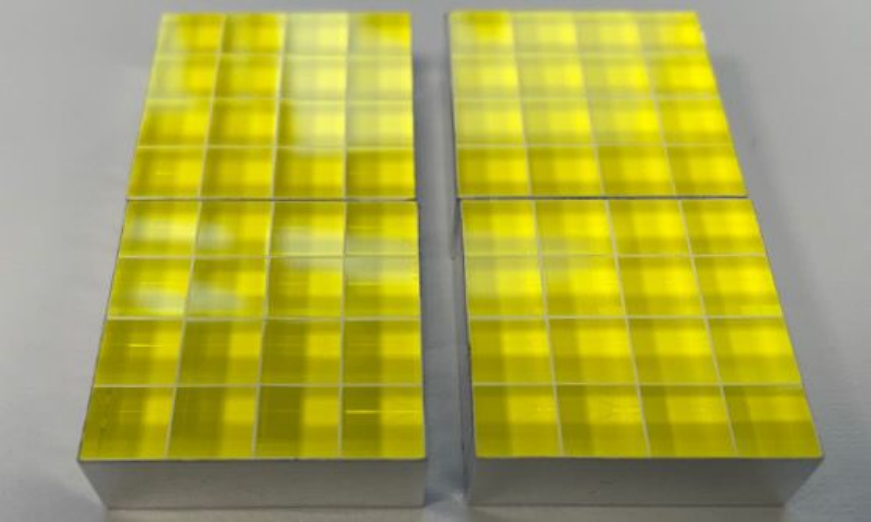}
        \caption{}
        \label{fig:2.2}
    \end{subfigure}
    \par
    \vspace{2mm}
    \begin{subfigure}{.48\linewidth}
        \centering
        \includegraphics[width=.95\linewidth]{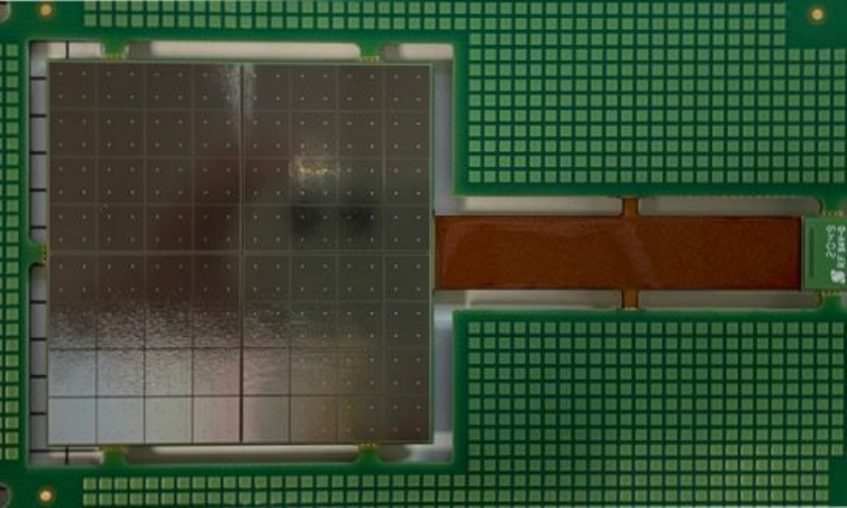}
        \caption{}
        \label{fig:2.3}
    \end{subfigure}%
    \begin{subfigure}{.48\linewidth}
        \centering
        \includegraphics[width=.95\linewidth]{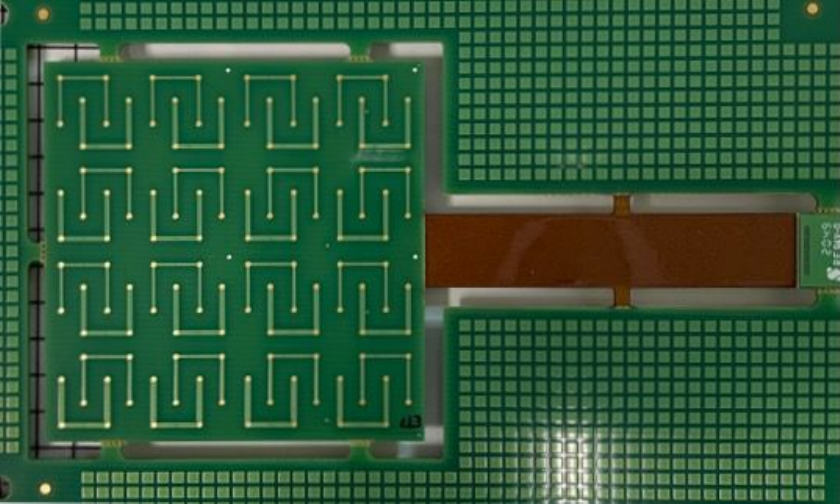}
        \caption{}
        \label{fig:2.4}
    \end{subfigure}
    \caption{A GTM detector and its components. (a) The scintillator arrays shown in (b) are upside down to couple with the SiPMs arrays on PCB shown in (c) using BC-630 to form a unit. (b) 4 pieces of $4 \times 4$ \ce{GAGG(Ce)} arrays. (c) 4 pieces of $4 \times 4$ SiPMs arrays on PCB. (d) PCB with 16 combined readout channels.}
    \label{fig:2}
\end{figure}

\cref{fig:2} decomposes the internal composition of a detector. When the outer aluminum shell and the thermal pad (about 0.38 mm thick) are removed, a detector consists of three objects, \cref{fig:2.1}. The first object is made up of 4 pixelated \ce{GAGG(Ce)} scintillator arrays, \cref{fig:2.2}. For each array, except for the face in contact with the second object, the remaining five faces are wrapped with a layer of aluminum film (0.05 mm thick). Within the film, there are 16 \ce{GAGG(Ce)} bars (each $6\ {\rm mm} \times 6\ {\rm mm} \times 8\ {\rm mm}$) arranged in a $4 \times 4$ configuration, forming 16 pixels (each $6\ {\rm mm} \times 6\ {\rm mm}$). The five faces of each \ce{GAGG(Ce)} bar, aligned with the covering film, are surrounded by a \ce{BaSO4} layer (0.2 mm thick) for reflection. At the 520 nm scintillation emission wavelength of \ce{GAGG(Ce)} \citep{Yoneyama_2018}, the reflectance of \ce{BaSO4} is nearly $100\ \%$ \citep{Knighton_2005}. The second object is composed of 4 multi-channel SiPMs arrays (Hamamatsu S13361-6050NE-04), \cref{fig:2.3}. The channel and gap dimensions of SiPMs in an array are consistent with the \ce{GAGG(Ce)} pixel size and \ce{BaSO4} thickness. With this size consistency, the side without \ce{BaSO4} and aluminum film of \ce{GAGG(Ce)} can be coupled with the SiPM using optical silicone grease (Saint Gobain BC-630, a few tens of microns thick). The third object is the Printed Circuit Board (PCB) used to read out the signal from SiPMs, \cref{fig:2.4}. Considering the power constraint of 2 W for the orbital average, signals must be collected from the adjacent $2 \times 2$ SiPMs and treated as a single readout channel (each about $12\ {\rm mm} \times 12\ {\rm mm}$). For more details about GTM's mechanical structure design, please refer to a forthcoming GTM paper.

\subsection{Readout Electronics}
\label{sec:2.2}

To manipulate 64 channels in a module, we use 2 Application-Specific Integrated Circuit (ASIC) chips, known as Cherenkov Imaging Telescope Integrated Read Out Chip (CITIROC, Weeroc CITIROC 1A) \citep{Perri_2023}. One advantage of CITIROC is its ability to apply a uniform High Voltage (HV) to 32 channels while also permitting individual fine-tuning of gain and threshold through the application of a low voltage (0 - 4.5 V) to each channel via the Digital-to-Analog Converter (DAC) \citep{Impiombato_2015}. In addition, CITIROC offers High Gain (HG) and Low Gain (LG) preamplifiers to amplify the analog signal of each channel, facilitating a broad energy range and high resolution at low energy. \cref{sec:3} explains more details on our HV, DAC and LG/HG settings. Those analog signals are finally converted into digital signals by the Analog-to-Digital Converter (ADC) and sent to the Field Programmable Gate Array (FPGA) in the master module. Altogether, 4 CITIROCs are used to read out 128 channels in GTM. \cref{tab:1} lists the mapping of the detectors to the readout system and their respective positions on the satellite. 

\begin{table}[t]
    \centering
    \begin{threeparttable}
        \caption{The electronic readout channels of 8 GTM detectors and their corresponding pointing within the satellite coordinate.}
        \label{tab:1}
        \begin{tabular*}{1\linewidth}{@{\extracolsep{\fill}}ccccc}
            \toprule
                Module & Detector & CITIROC & Channel & Orientation\tnote{a} \\
            \midrule
                \multirow{4}{*}{M} & 1 & \multirow{2}{*}{1} & 16-31 & PP \\
                                   & 2 &                    & 0-15  & PT \\
                                   & 3 & \multirow{2}{*}{0} & 16-31 & PB \\
                                   & 4 &                    & 0-15  & PN \\
            \cmidrule(lr){1-5}
                \multirow{4}{*}{S} & 1 & \multirow{2}{*}{1} & 16-31 & NN \\
                                   & 2 &                    & 0-15  & NT \\
                                   & 3 & \multirow{2}{*}{0} & 16-31 & NB \\
                                   & 4 &                    & 0-15  & NP \\
            \bottomrule
        \end{tabular*}
        \begin{tablenotes}
            \item[a]{The name defined in coordinate of FS-8B \citep{Chang_2022}.}
        \end{tablenotes}
  \end{threeparttable}
\end{table}

\begin{figure}[t]
    \centering
    \includegraphics[width=.9\linewidth]{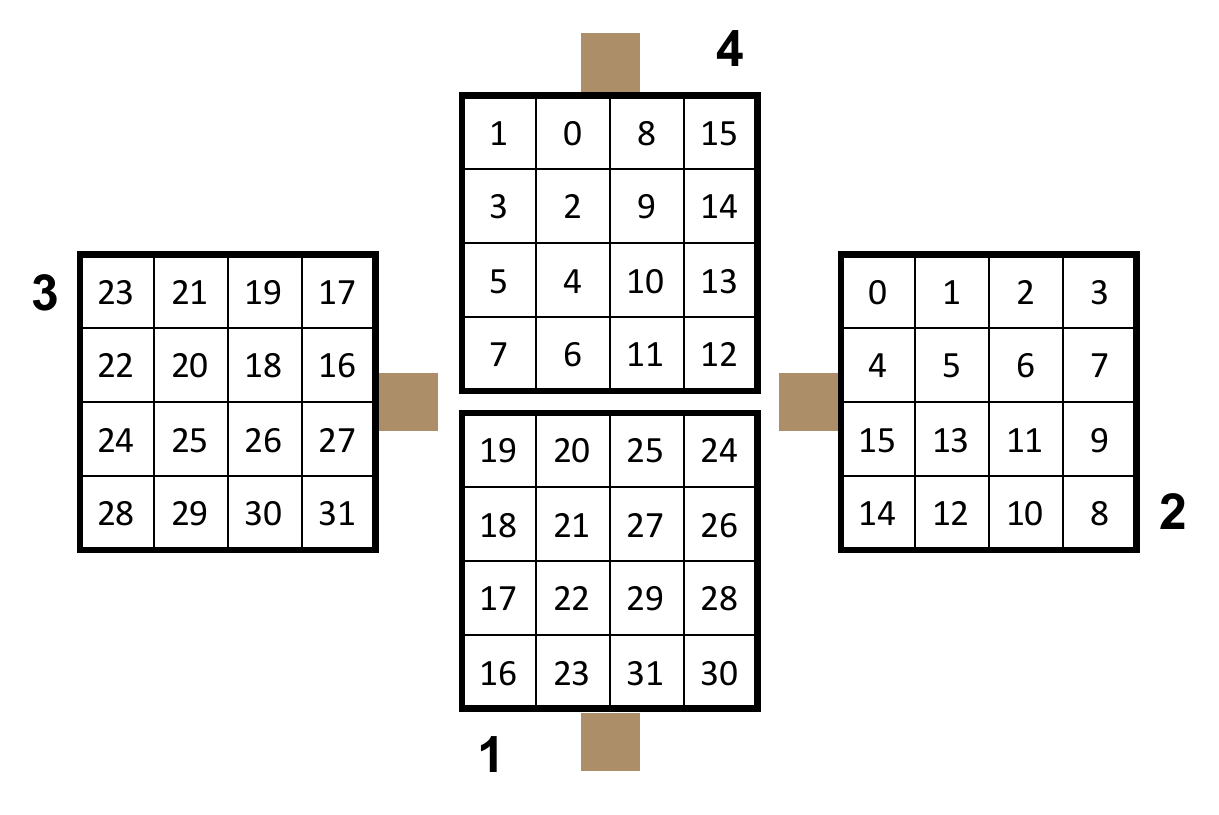}
    \caption{The projection of a GTM module. 4 detectors are labeled in bold. The numbers in the $4 \times 4$ grid correspond to the channels in \cref{tab:1}. The brown color indicates the direction of the readout wire.}
    \label{fig:3}
\end{figure}

For an illustration of how the channels in \cref{tab:1} are distributed in \cref{fig:1}, a schematic of \cref{fig:3} is created by elevating the edge of the detector closest to the ground to the same height as the edge farthest from the ground. This information is essential for data analysis, as it allows for the verification of data integrity and the potential reconstruction of multiple hit events produced by Compton scattering.

\subsection{GRB Saturation Fluence}
\label{sec:2.3}

The photon-counting data of GTM has a minimum time resolution of about $3.8\ {\rm \mu s}$ due to 18-bit dynamic memory used to record the fine time counter in FPGA. Although the decay time for \ce{GAGG(Ce)} and the time constant for SiPM are fast enough (about 100 ns and in the ps level, respectively), a CITIROC takes about $22.3\ {\rm \mu s}$ to process a triggered signal, resulting in an ideal saturation rate about 44.9 kHz. When the system is busy, the effect of buffer time must be taken into account. Nevertheless, here we use this ideal saturation rate to estimate the ideal GRB saturation fluence.

MEGAlib was used to simulate the background and source count rates. The simulation considered a 15 keV threshold and only selects events with deposition energy between 50 keV and 1 MeV. The background simulation included input energy from 1 keV to 1 TeV and spectra containing many albedo and cosmic components. The background count rate for each detector is approximately 50 Hz. The source simulation used input energy ranging from 10 keV to 10 MeV and a spectrum that follows a cutoff power law (the photon flux density in units of ${\rm [\# / (s \times cm^2 \times keV)]}$). For SGRB, the power index is -0.5, with a peak energy of 500 keV, a duration of 0.5 s, and a fluence of $2 \times 10^{-4}\ {\rm erg/cm^2}$. Conversely, LGRB has a power index of -1, a peak energy of 300 keV, a duration of 10 s, and a fluence of $4 \times 10^{-4}\ {\rm erg/cm^2}$. GRBs were arbitrarily incident from $\theta=135\ ^\circ$ and $\phi=60$, $120$, $240$, and $300\ ^\circ$ in FS-8B's coordinates, which gave more counts for the PT+PP, PT+PN, NT+NN, and NT+NP detectors, respectively. To estimate the saturation fluence of GRB, it is logical to choose the incident direction that yields more counts for 2 detectors controlled by the same CITIROC, e.g., PT+PP and NT+NN. For this type of CITIROC, the source count rates for SGRB and LGRB are approximately 28.6 and 3.5 kHz, respectively. It can be calculated that the ideal saturation fluences of SGRB and LGRB are about $3.1 \times 10^{-4}$ and $5.0 \times 10^{-3}\ {\rm erg/cm^2}$, respectively.

\section{Measurements}
\label{sec:3}

\subsection{Threshold Setup}
\label{sec:3.1}

Even though the aluminum shell around the detector blocks most low-energy photons, the system may still generate low-energy noise during operation. To prevent this noise from causing anomalies and overwhelming the GTM, the DAC controlling the threshold needs to be set. We measured the background signal in ambient conditions and gradually raised the threshold until reaching a specific DAC value. At that DAC value, we observed a significant decrease in the background count rate. Therefore, this particular DAC value was set as the threshold. Indeed, the threshold should be fine-tuned for different readout channels, environmental temperatures, and operating voltages. To minimize complexity, a uniform threshold setting was applied to all readout channels for all subsequent measurements.

\subsection{Isotope Calibration}
\label{sec:3.2}

To correctly transform the measured ADC back to energy, we can only rely on isotopes with well-known peaks to label the corresponding ADC values and thus establish the mapping. Furthermore, in the case of real measurements, those peaks are not delta functions. They most likely form a Gaussian distribution due to uncertainties. This means that when we irradiate GTM with numerous radioactive sources, we can also analyze the width of the Gaussian distribution to determine the energy resolutions of GTM detectors.

\begin{figure}[t]
    \centering
    \includegraphics[width=.8\linewidth]{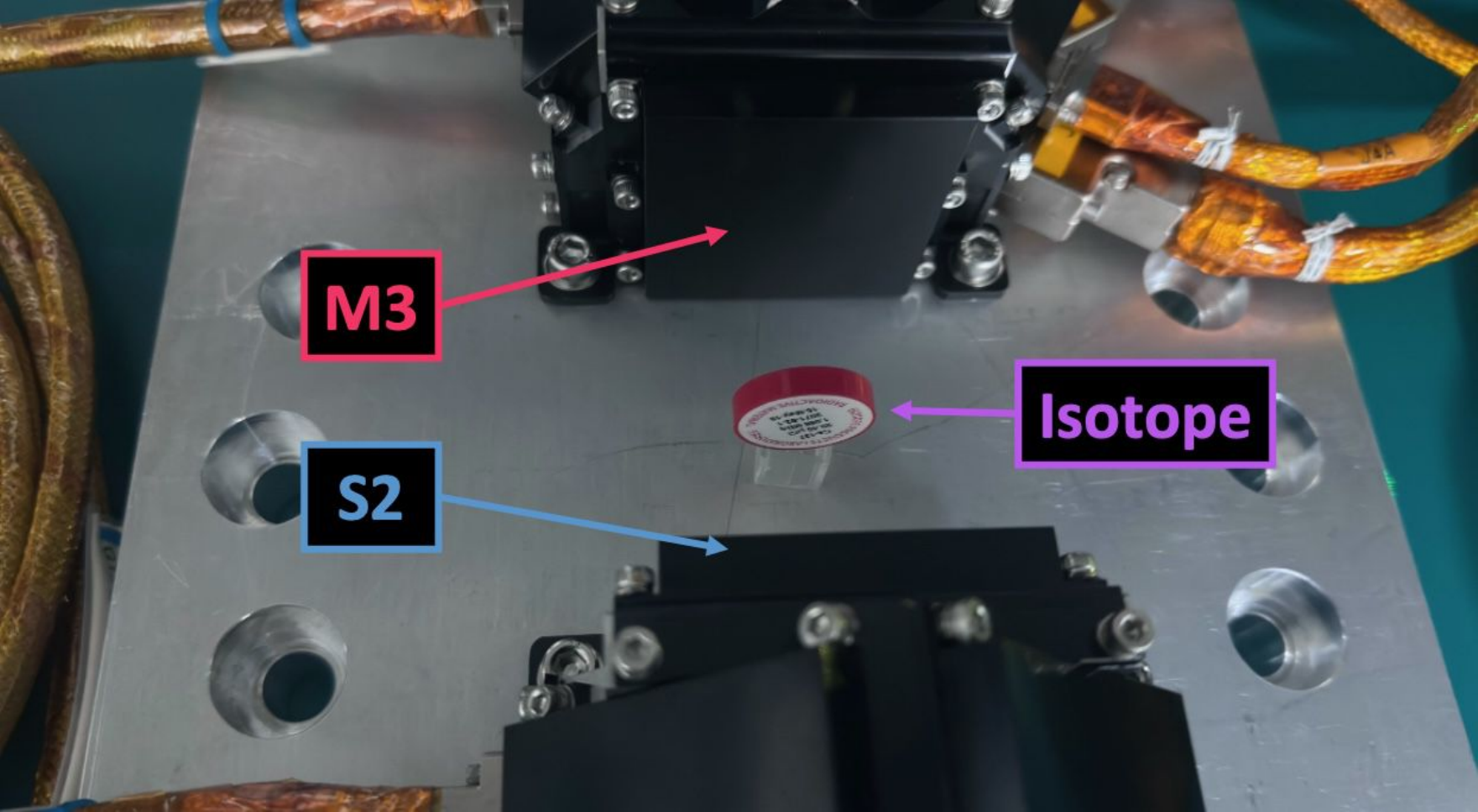}
    \caption{The setup of the isotope calibration. Except for \ce{LYSO(Ce)}, the isotope is placed at the middle of GTM's M3 and S2 detectors.}
    \label{fig:4}
\end{figure}

\cref{fig:4} shows the setup of measurements performed in ambient conditions (board temperature about $30\ {\rm ^\circ C}$) at Academia Sinica in Taiwan. The sources we used are listed in \cref{tab:2} \citep{Bissaldi_2009}. \cref{sec:3.3} will explain more about measurements using Lutetium Yttrium Orthosilicate doped with Cerium [\ce{LYSO(Ce)}]. It is important to note that the amplification of each readout channel differs even when the same HV and DAC are applied. Through some trials, we adjusted the DAC at a fixed HV to ensure that the maximum energy measured for each readout channel is approximately 1 MeV at LG/HG = 2/20. If we need to observe isotopes with peak values greater than 1 MeV, we set LG/HG = 1/10. The function of HG is to obtain better resolution in the relatively low-energy region. The results are displayed in \cref{sec:5.1}.

\begin{table}[t]
    \centering
    \begin{threeparttable}
        \caption{Measurements in ambient conditions with different radioactive sources and LG/HG settings to study energy resolution.}
        \label{tab:2}
        \begin{tabular*}{1\linewidth}{@{\extracolsep{\fill}}ccccc}
             \toprule
                Source & Peak [keV]\tnote{a} & LG/HG & HV [V]\tnote{b} & DAC\tnote{c} \\
            \midrule
                \ce{^{133}Ba} & 81 $\&$ 356    & 2/20 & 55  & certain \\
                \ce{^{57}Co}  & 122            & 2/20 & 55  & certain \\
                \ce{LYSO(Ce)}     & 202 $\&$ 307   & both & 55  & certain \\
                \ce{^{22}Na}  & 511 $\&$ 1274  & 1/10 & 55 & certain \\
                \ce{^{137}Cs} & 662            & both & 55  & certain \\
                \ce{^{60}Co}  & 1173 $\&$ 1332 & 1/10 & 55 & certain \\
            \bottomrule
        \end{tabular*}
        \begin{tablenotes}
            \item[a]{The peak values are used to calculate energy resolution later.}
            \item[b]{55: (55.8, 55.9, 55.8, 55.6) for (M CITIROC 0, M CITIROC 1, S CITIROC 0, S CITIROC 1).}
            \item[c]{certain: Different DACs are set for all 128 readout channels to make their amplification roughly the same.}
        \end{tablenotes}
    \end{threeparttable}
\end{table}

\subsection{Gain Calibration}
\label{sec:3.3}

\begin{figure}[t]
    \centering
    \begin{subfigure}{.48\linewidth}
        \centering
        \includegraphics[width=1\linewidth]{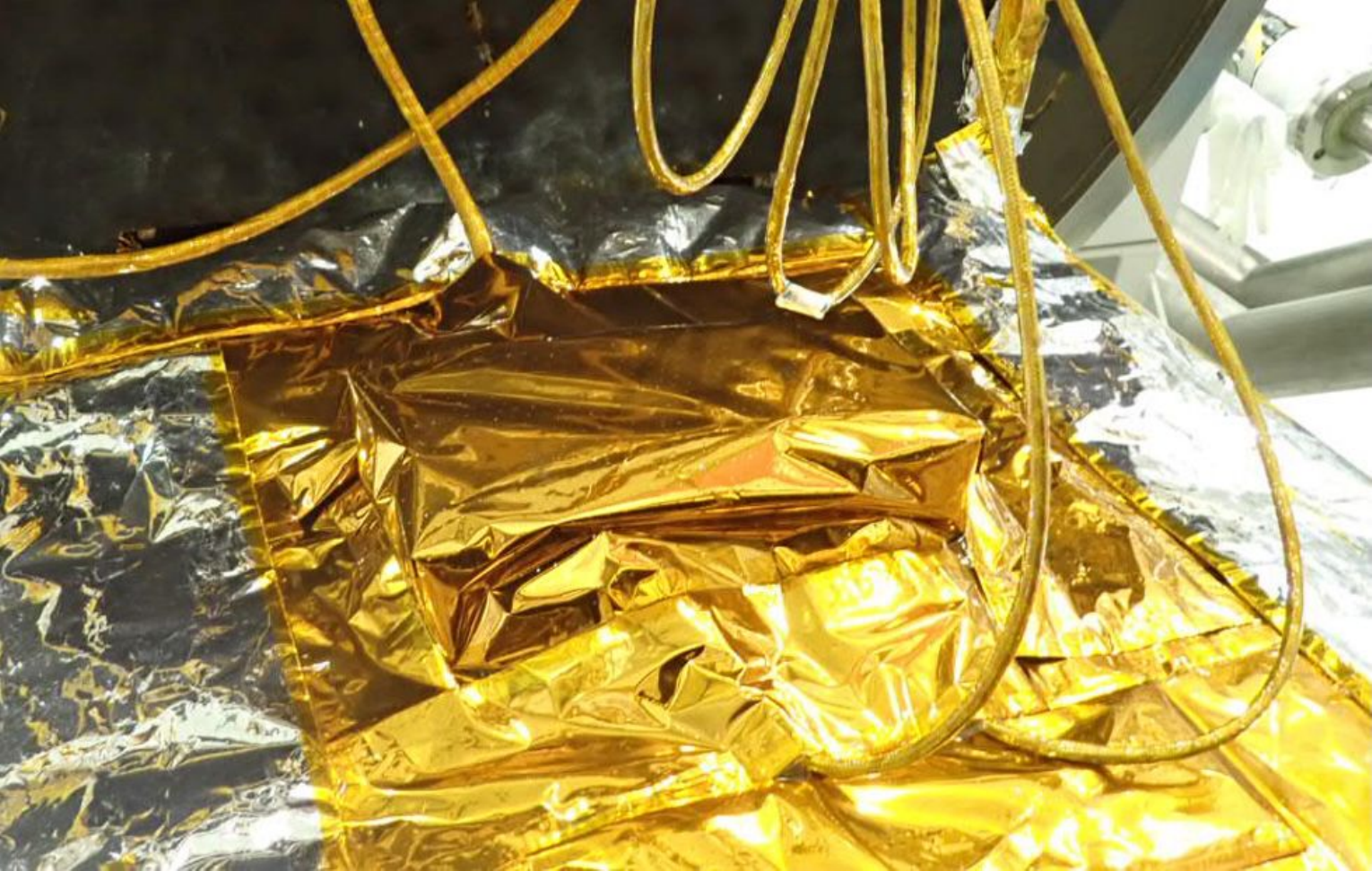}
        \caption{}
        \label{fig:5.1}
    \end{subfigure}
    \begin{subfigure}{.48\linewidth}
        \centering
        \includegraphics[width=1\linewidth]{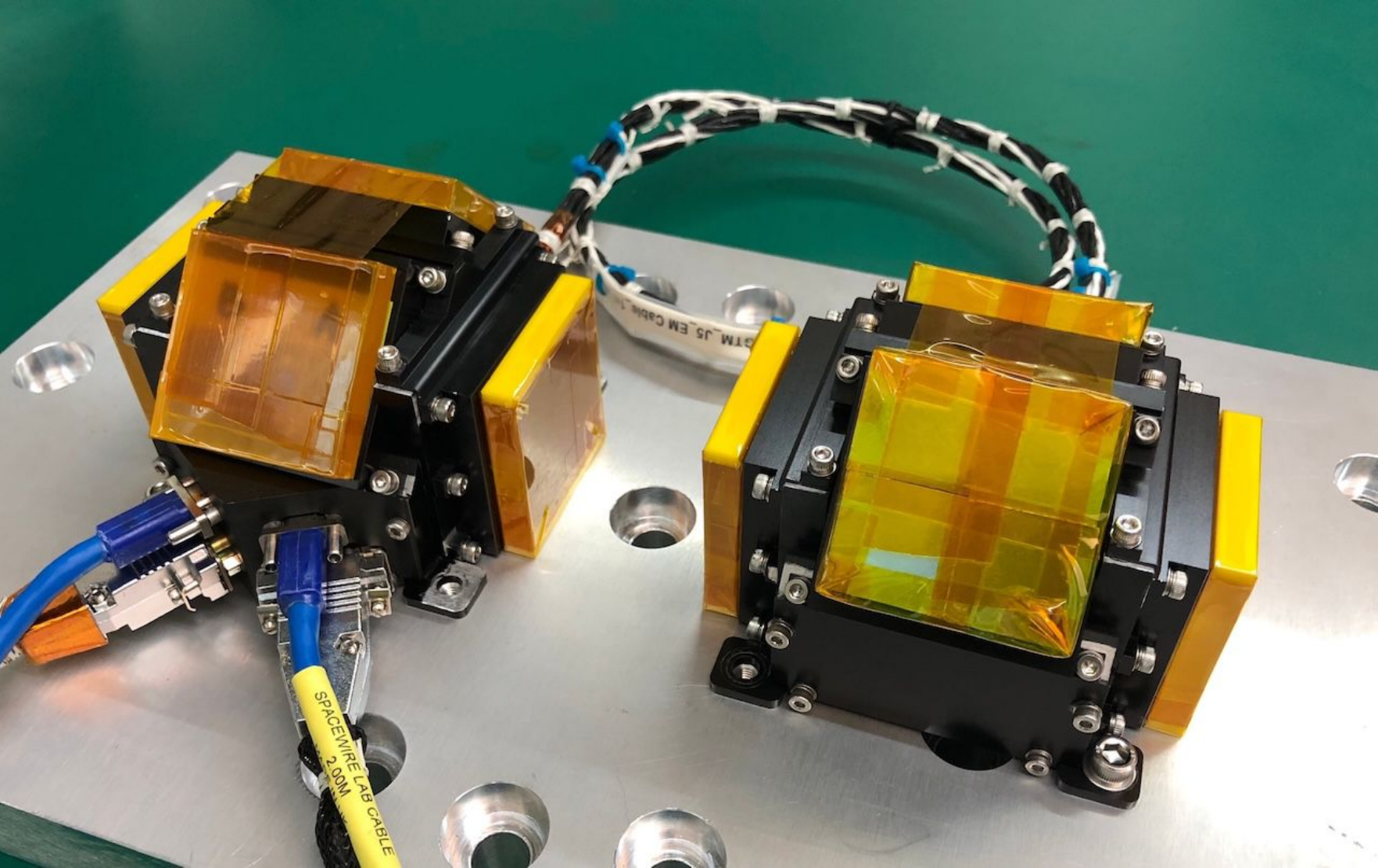}
        \caption{}
        \label{fig:5.2}
    \end{subfigure}
    \caption{The setup in the TV chamber. (a) GTM is covered by Multi-Layer Insulation (MLI) to mimic the real operations in space. (b) 8 \ce{LYSO(Ce)} crystals are wrapped in high-temperature thermal tape and secured to 8 GTM detectors.}
    \label{fig:5}
\end{figure}

The conversion between incident energy and readout signal is not constant. Instead, this relationship is a function of both temperature and voltage. To understand how this relationship evolves, we conducted a series of measurements in the Thermal-Vacuum (TV) chamber at TASA, \cref{fig:5.1}. The chamber can create a space-like vacuum and vary the environmental temperature between $-20$ and $70\ {\rm ^\circ C}$ to simulate the conditions in orbit. Unfortunately, carrying and changing isotopes within the TV chamber is challenging. With the isotope calibration discussed in \cref{sec:3.2}, we observed that the X-Ray Fluorescence (XRF) and intrinsic emission of \ce{LYSO(Ce)} exhibit distinct peaks at 55 keV ($K_{\alpha}$ XRF of \ce{^{176}Lu}) \citep{Eriksson_2003} and at 202 and 307 keV ($\gamma$-ray emissions due to the $\beta$ decay of \ce{^{176}Lu}) \citep{Alva_2018}. These peaks provide a suitable non-regulated source for labeling. \cref{fig:5.2} shows how we placed \ce{LYSO(Ce)} crystals (each about $50\ {\rm mm} \times 50\ {\rm mm} \times 8\ {\rm mm}$) to irradiate GTM in TV chamber. \cref{tab:3} arranges all the analyzable measurements made in the TV chamber. The results are displayed in \cref{sec:5.2,sec:5.3}.

\begin{table}[t]
    \centering
    \begin{threeparttable}
        \caption{Measurements in the TV chamber with \ce{LYSO(Ce)} at various temperatures and voltages to study gain evolution.}
        \label{tab:3}
        \begin{tabular*}{1\linewidth}{@{\extracolsep{\fill}}cccc}
            \toprule
                {Temperature [${\rm ^\circ C}$]}\tnote{a} & LG/HG & HV [V]\tnote{b} & DAC\tnote{c} \\
            \midrule
                -10 & 2/20 & 55 & certain \\
                -5  & 2/20 & 55 & certain \\
                0   & 2/20 & 55 & certain \\
                0   & 2/20 & 55 & 0x01 \\
                0   & 2/20 & 55 & 0x03 \\
                0   & 2/20 & 56 & 0x01 \\
                0   & 2/20 & 57 & 0x01 \\
                5   & 2/20 & 55 & certain \\
                10  & 2/20 & 55 & certain \\
                20  & 2/20 & 55 & certain \\
            \bottomrule
        \end{tabular*}
        \begin{tablenotes}
            \item[a]{-10: (-1, -3), -5: (3, 2), 0: (8, 7), 5: (13, 12), 10: (18, 17), and 20: (27, 26) for (M, S) due to waste heat from the operation.}
            \item[b]{56: (56.4, 56.5, 56.4, 56.2) and 57: (57.1, 57.1, 57.0, 56.8) for (M CITIROC 0, M CITIROC 1, S CITIROC 0, S CITIROC 1).}
            \item[c]{Hexadecimal → Binary → Reversed Binary → Decimal to represent a ratio (Decimal/255) to control 0 - 4.5 V.}
        \end{tablenotes}
    \end{threeparttable}
\end{table}

\section{Data Processing}
\label{sec:4}

\subsection{Raw Data Visualization}
\label{sec:4.1}

\begin{figure}[t]
    \centering
    \includegraphics[width=.9\linewidth]{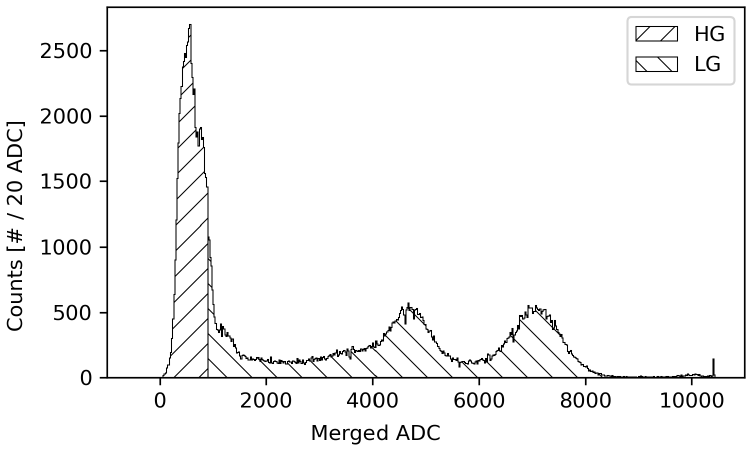}
    \caption{The ADC spectrum of M3 channel 18. It shows all hits regardless of whether they are from single-hit or multiple-hit events. The label of the x-axis is called 'Merged ADC' since the HG data (filled with slash lines) is transformed to the scale of the LG data (filled with backslash lines).}
    \label{fig:6}
\end{figure}

To clearly illustrate how we executed data reduction on raw data and what we uncovered during the analysis process, let's examine the $-10\ {\rm ^\circ C}$ measurement referenced in \cref{tab:3}. We will focus on the M3 detector, which was tested in both isotope and gain calibration measurements, for demonstration. After decoding the raw binary data from GTM, the scientific data follows a photon-counting format. In each event data, all triggered hits are kept. The information for a single hit includes its triggered module, CITIROC, channel, preamplifier, and ADC. The ADC range of -5,383 to 11,000 is recorded using 14-bit dynamic memory. Besides, data amplified by HG will be end at ADC = 8,192, while the remaining data will be amplified by LG and terminated at about ADC = 10,500. To concatenate HG and LG data for analysis, HG data can ideally be divided by the HG/LG ratio to merge it back to the LG scale. Nevertheless, in reality, all ratios for all channels can vary owing to voltage fluctuations. As a result, the low-ADC edge of LG data must first be found, and then the real ratio is calculated as 8,192 divided by that edge value. \cref{fig:6} plots all ADCs fired in channel 18 of M3 and has merged HG into LG data. In the figure, a tiny amount of data is omitted (about $0.082\ \%$), primarily from the HG, which forms a bump near ADC = 0, away from the threshold drop, because this data comes from the noise after the threshold excision. It is obvious that the 202 and 307 keV peaks from \ce{LYSO(Ce)} are visible between 4,000 to 8,000 ADC.

\begin{figure}[t]
    \centering
    \includegraphics[width=1\linewidth]{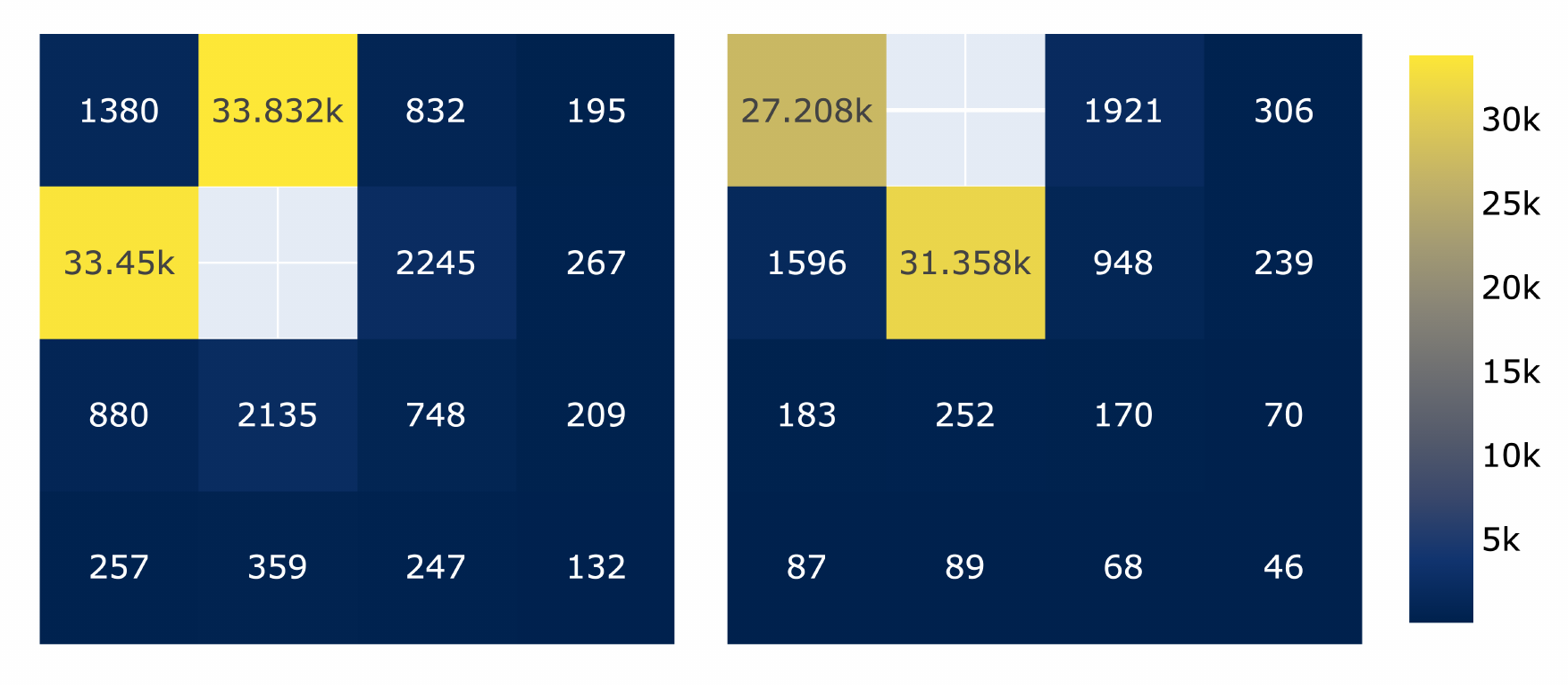}
    \caption{The spatial distribution of all residual hits for the highest energy hit in channels 18 (left) and 16 (right) of M3. The vacancy indicates the channel with the highest energy hit, while the numbers labeled on all the other channels represent the accumulated counts of their appearances.}
    \label{fig:7}
\end{figure}

For the origin of the enormous number of hits between 0 and 2,000 ADC, we initially naively regarded them mainly as the low-energy contribution from multiple-hit events due to Compton scattering. However, when we selected only the single-hit events to plot the ADC spectrum similar to that shown in \cref{fig:6} to see the 55 keV peak from \ce{LYSO(Ce)}, we surprisingly noticed that the peaks at 202 and 307 keV became too small or even disappeared in all channels despite the appearance of the 55 keV peak. To understand this unusual phenomenon, we created the spatial distribution of multiple-hit events, as shown in \cref{fig:7}, which reveals unnatural patterns. To obtain this spatial distribution, a rough linear ADC-to-energy relation was fitted using the 202 and 307 keV peaks in \cref{fig:6}. By this relation, the highest energy hit and the other hits in a multiple-hit event can be separated. This allows us to extract all multiple-hit events with the highest energy hit in a given channel and count the appearance of their residual hits in the remaining channels in the same detector.

In view of 3D Compton scattering, the scattered photons are independent of the azimuthal angle if the direction of the incident photons without polarization is treated as the zenith. That is to say, the accumulated counts of residual hits in the adjacent channels should statistically be at a similar level. \cref{fig:7} shows that the patterns are inconsistent with this expectation. It implies that an energetic hit tends to create hits in two specific nearest directions. By comparing these patterns with the inner structure of the detector shown in \cref{fig:2}, it seems that the two unique directions correlate with the gaps between the four pieces of \ce{GAGG(Ce)} and also SiPMs arrays. The scenario suggests that channel noise exists in the data due to crosstalk, and the physical gap between all pieces of \ce{GAGG(Ce)} and SiPM arrays reduces the probability of detecting the noise.

\subsection{Crosstalk Correction}
\label{sec:4.2}

To address the crosstalk issue, a two-step correction was applied to mitigate the negative impact of channel noise. First, residual hits in the surrounding circle of the channel with the highest energy hit were treated as noise. After removing these, the event ratio of single-hit events increased from $36$ to $95\ \%$. As a result, we were able to simultaneously observe the 55, 202, and 307 keV peaks in the ADC spectrum using only single-hit events, which allowed for a better linearity fit of the ADC-to-energy relation. However, this approach may mistakenly classify actual multiple hits in adjacent channels as noise. Therefore, a criterion must be established to distinguish genuine Compton scattering hits from noise.

\begin{figure}[t]
    \centering
    \includegraphics[width=.9\linewidth]{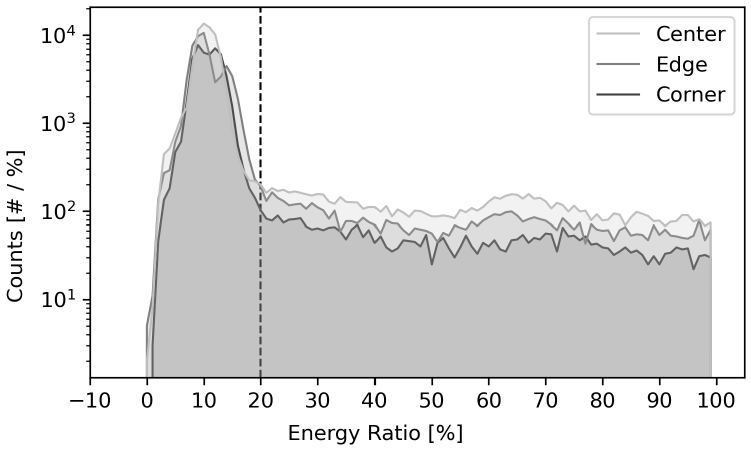}
    \caption{The energy ratio distribution of residual hits in the surrounding circle for the highest energy hit in channels 18, 16, and 17 of M3. These 3 channels correspond to positions at the center, edge, and corner of the detector, respectively. The vertical dashed line marks the position of the $20\ \%$ ratio, and hits below this ratio are considered noise.}
    \label{fig:8}
\end{figure}

In the next step, the improved ADC-to-energy relation was used to reprocess the data, redefining each event’s highest energy hit and its associated residual hits. By collecting the relative energy ratio of each residual hit with respect to the corresponding highest energy hit, \cref{fig:8} was created. In this figure, we present 3 cases to demonstrate different morphological types due to the system's symmetrical design. Based on the results from all measurements, we ultimately selected a cutoff at $20\ \%$, considering residual hits with an energy ratio smaller than $20\ \%$ as noise. After applying the second correction, the ratio of single-hit events decreased from $95$ to $87\ \%$. This not only preserved a sufficient number of single-hit events for fitting the final ADC-to-energy relation, but also provided more multiple-hit events, enhancing performance in the high energy region.

\begin{figure}[t]
    \centering
    \includegraphics[width=.9\linewidth]{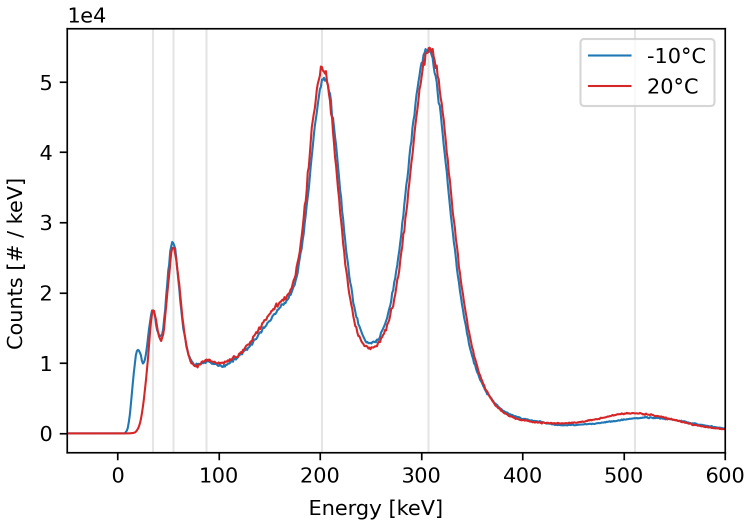}
    \caption{The corrected GTM energy spectrum of the $-10$ (blue) and $20\ {\rm ^\circ C}$ (red) measurements. The gray vertical lines label the 35, 55, 88, 202, 307, and 511 keV from left to right.}
    \label{fig:9}
\end{figure}

By following the procedure described above, all measurements listed in \cref{tab:3} can be recovered. Similarly, to heal the spectrum of the measurements arranged in \cref{tab:2}, the precise ADC-to-energy relation used to indicate noise can be fitted using peaks of isotopes. \cref{fig:9} plots the GTM energy spectrum of measurements taken at $-10$ and $20\ {\rm ^\circ C}$ after correcting the crosstalk. Both spectra display not merely the strong 55, 202 and 307 keV peaks, but also the weak 88 keV peak ($\gamma$-ray emission due to $\beta$ decay of \ce{^{176}Lu} with small escaping probability). In the measurement at $-10\ {\rm ^\circ C}$, the higher gain caused by the lower temperature, compared to the $20\ {\rm ^\circ C}$ measurement, is equivalent to cutting the threshold at a lower energy level. This confirms the detection of the 35 keV peak ($K_{\alpha}$ and $K_{\beta}$ XRF of Ba in \ce{BaSO4} \citep{Cho_2014,Durdu_2022}), which might serve as a future onboard calibration peak. However, the high gain in the $-10\ {\rm ^\circ C}$ measurement also has its drawbacks. When observing the 511 keV annihilation line, the signal is distorted due to its proximity to the unstable gain region near the ADC termination. As a result, in the $-10\ {\rm ^\circ C}$ measurement, all events with hit $>$ 10,000 ADC are ignored. With this adjustment, a false peak around 400 keV can be avoided, but the impact of missing single-hit events (whose effective area decreases with increasing energy) still remains, making its significance and accuracy inferior to that of the $20\ {\rm ^\circ C}$ measurement.

\section{Results and Discussion}
\label{sec:5}

\subsection{Energy Resolution}
\label{sec:5.1}

In gain calibration, the linear ADC-to-energy relation can be effectively fitted using 3 relatively low-energy peaks. However, in isotope calibration, the highest peak is over 1 MeV. For \ce{GAGG(Ce)}, the light yield is about 56,000 photons at 1 MeV. The quantum efficiency of the SiPM at 520 nm is around $40\ \%$. The expected number of scintillating photons a SiPM pixel should detect is approximately 22,400 photons, which exceeds the 14,336 SPADs in a SiPM pixel. Therefore, the SiPM's saturation effect should be taken into account. The real fired number of SPADs ($N_{f}$), considering only the Leading Order (LO) correction, is described by \cref{eq:1} \citep{kotera_2015}.
\begin{equation}
    \label{eq:1}
    N_{f}^{LO} = N_{S} \cdot (1 - {\rm e}^{- \epsilon \cdot N_{in} / N_{S}})
\end{equation} where $N_{S}$ is the number of SPADs in a SiPM pixel, $\epsilon$ is the detection efficiency of the SiPM, and $N_{in}$ is the number of incident scintillating photons on a SiPM pixel.

By multiplying the gain of a single SPAD ($G_{S}$) with all the numbers $N$ that appear in \cref{eq:1}, and considering that $N_{in}$ is actually proportional to the incident energy ($E$), we can rewrite the expression using the fitting parameters $\alpha$ and $\beta$, as shown in \cref{eq:2}.
\begin{equation}
    \label{eq:2}
    ADC = \alpha \cdot (1 - {\rm e}^{- \beta \cdot E / \alpha})
\end{equation}

\begin{figure}[t]
    \centering
    \includegraphics[width=.9\linewidth]{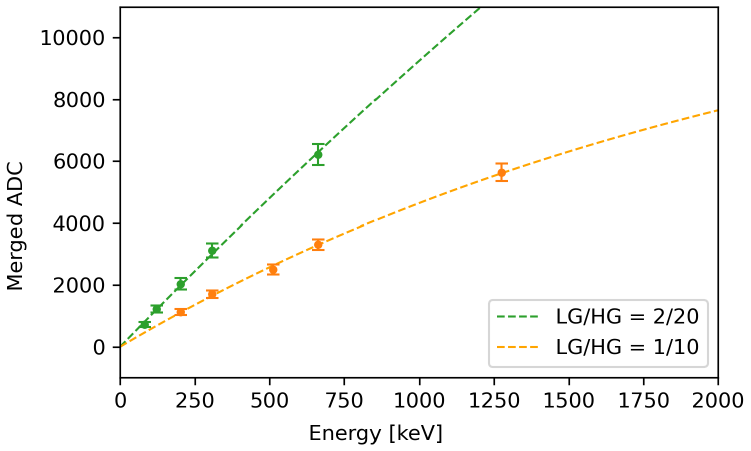}
    \caption{The energy-to-ADC mapping of M3 channel 17. The green and orange points mark the peak of isotopes measured at different LG/HG settings. To avoid mislabeling the ADC values corresponding to energy, 356 keV for \ce{^{133}Ba} and 1173 and 1332 keV for \ce{^{60}Co} are excluded from the fitting since they are more easily contaminated by nearby components. The dashed curves plot the fitting results with \cref{eq:2}.}
    \label{fig:10}
\end{figure}

\begin{figure*}[t]
    \centering
    \includegraphics[width=.95\linewidth]{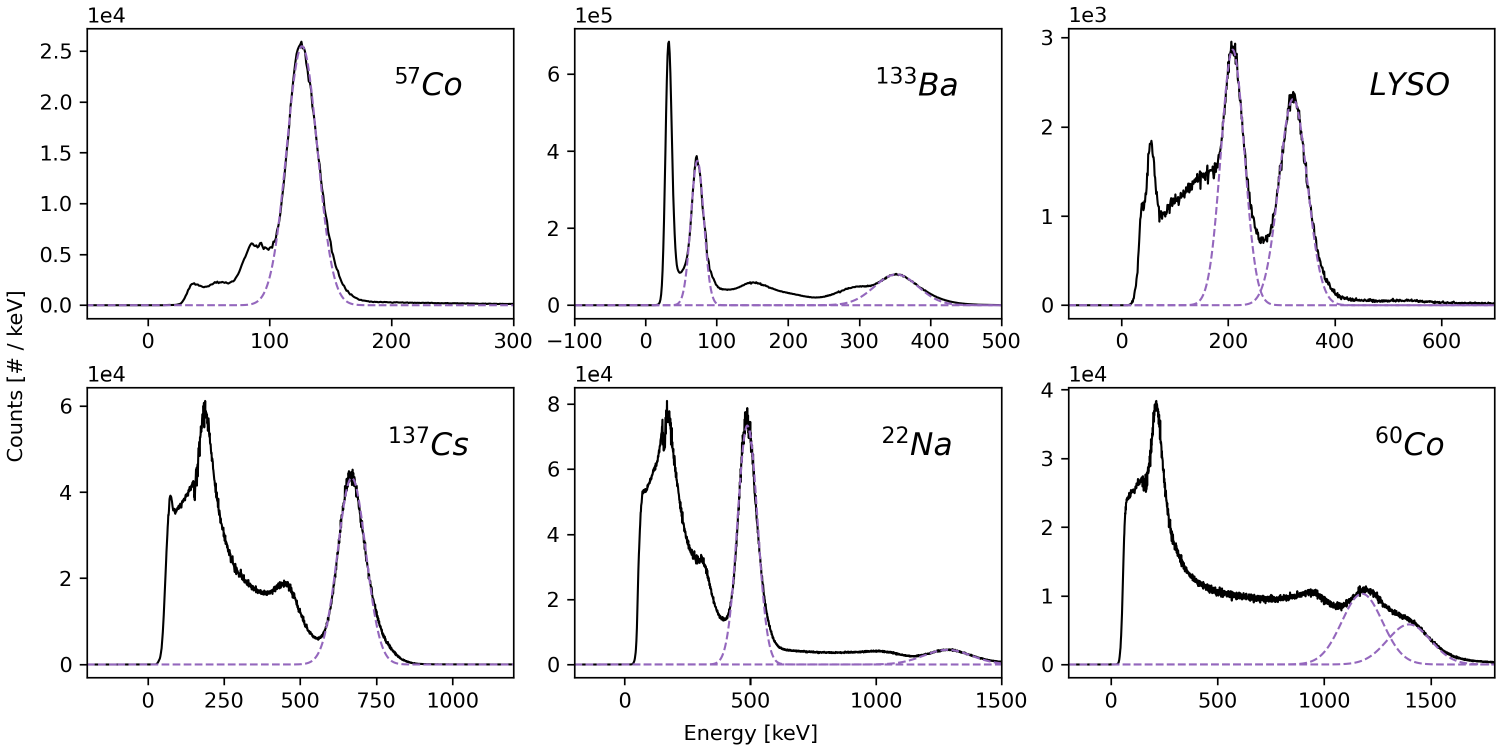}
    \caption{The energy spectrum of \ce{^{57}Co}, \ce{^{133}Ba}, \ce{LYSO(Ce)}, \ce{^{137}Cs}, \ce{^{22}Na}, and \ce{^{60}Co} measured by GTM M3 detector. The black solid curves plot the corrected spectrum considering the crosstalk correction mentioned in \cref{sec:4.2}. The purple dashed curves indicate the Gaussian fitting results at peak values listed in \cref{tab:2}.}
    \label{fig:11}
\end{figure*}

\begin{figure}[t]
    \centering
    \includegraphics[width=.9\linewidth]{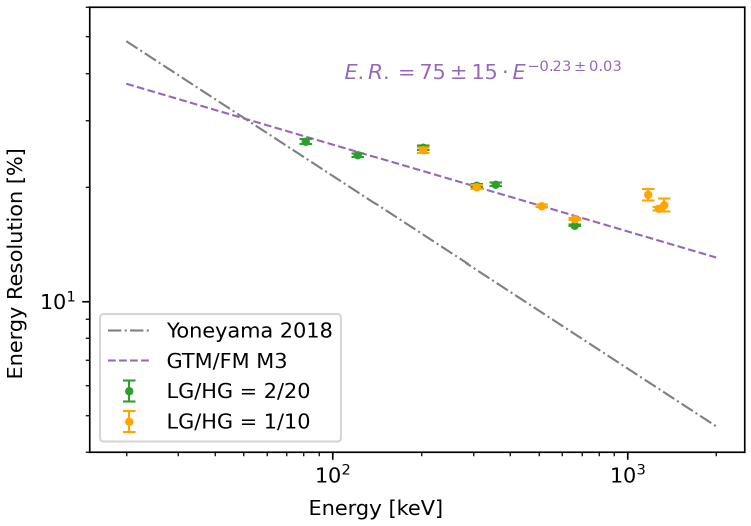}
    \caption{The energy resolution as a function of energy. The green and orange points represent measurements taken with different LG/HG settings. The purple dashed curve shows the best-fit line for the GTM M3 detector. The gray dashed curve is extracted from the literature \citep{Yoneyama_2018}, which shows the slope is about -0.5.}
    \label{fig:12}
\end{figure}

By means of \cref{eq:2}, all measurements listed in \cref{tab:2} can be successfully mapped to energy, accounting for the linear property in the lower energy region and the nonlinear effect in the higher energy domain simultaneously, as shown in \cref{fig:10}. \cref{fig:11} presents the comprehensive spectrum with the fitted Gaussian for all isotopes measured by the GTM M3 detector. From these fitted Gaussians, the energy resolution ($E.R.$) can be calculated using \cref{eq:3} \citep{Demir_2021}. In the denominator, we directly use the peak energy ($E_{0}$) mentioned in \cref{tab:2}. Both the value and the error for the Full Width at Half Maximum ($FWHM$) of $E_{0}$ can be obtained from \cref{eq:4}.
\begin{equation}
    \label{eq:3}
    E.R. = \frac{FWHM}{E_{0}} \times 100\%
\end{equation}
\begin{equation}
    \label{eq:4}
    FWHM = 2 \sqrt{2 {\rm ln} 2}\ \sigma
\end{equation} where $\sigma$ is the standard deviation of Gaussian distribution.

\cref{fig:12} illustrates the energy resolution comparison between GTM and another study that used a $5\ {\rm mm} \times 5\ {\rm mm} \times 5\ {\rm mm}$ \ce{GAGG(Ce)} crystal \citep{Yoneyama_2018}. It demonstrates that GTM has worse resolution. This may be due to the crosstalk issue and the fact that 4 SiPM pixels are combined into a single readout channel. At the 662 keV peak generated from \ce{^{137}Cs}, the energy resolution for GTM is approximately $16\ \%$, which is higher than the $7.6\ \%$ reported in the literature. For peaks larger than 1 MeV, the photoelectric effective area of \ce{GAGG(Ce)} becomes too small, and the Compton scattering cross section is further reduced by crosstalk correction, making the estimate of the $FWHM$ for these peaks more susceptible to the influence of nearby background features, which leads to poorer energy resolution.

\subsection{Voltage Dependence}
\label{sec:5.2}

\begin{figure}[t]
    \centering
    \begin{subfigure}{1\linewidth}
        \centering
        \includegraphics[width=.9\linewidth]{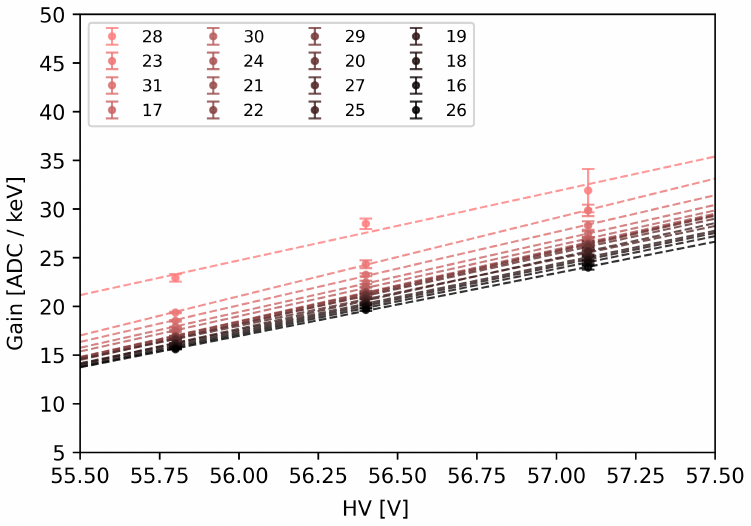}
        \caption{}
        \label{fig:13.1}
    \end{subfigure}
    \par
    \vspace{2mm}
    \begin{subfigure}{1\linewidth}
        \centering
        \includegraphics[width=.9\linewidth]{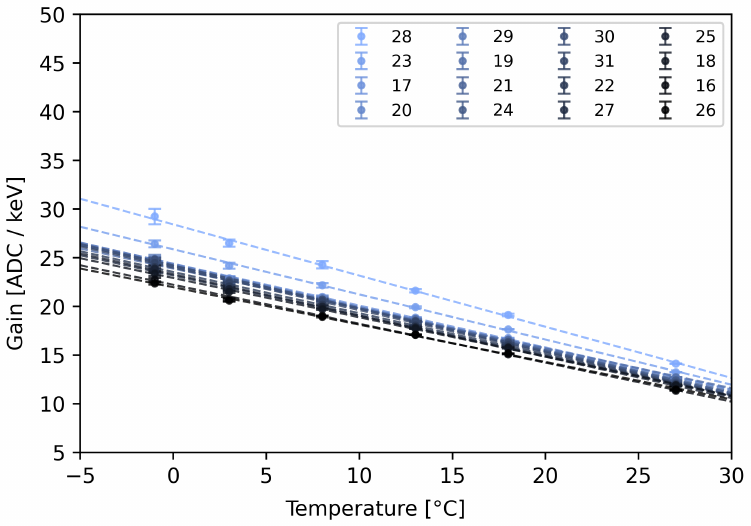}
        \caption{}
        \label{fig:13.2}
    \end{subfigure}
    \caption{The gain as a function of (a) HV (red) and (b) Temperature (blue) for 16 channels of GTM M3. The legend arranges the 16 channels in order of average gain (within the plotting range), from large (bright, upper left) to small (dark, lower right).}
    \label{fig:13}
\end{figure}

To avoid displaying too many similar figures, we select the GTM M3 detector (consistent with \cref{sec:5.1}) to illustrate how the gain changes with the applied overall HV in \cref{fig:13.1}. For the fine-tuned DAC, the linear relationship between gain and voltage still exists. The microstructure of the SiPM must be considered to apprehend why the gain variation of each channel is linearly proportional to the change in applied voltage. For a single SPAD, the gain is given by \cref{eq:5} \citep{Buzhan_2002}.
\begin{equation}
    \label{eq:5}
    G_{S} = \frac{Q_{S}}{e}
\end{equation} where $Q_{S}$ represents the accumulated charge during the avalanche process and $e$ denotes the charge of the electron excited by an incident photon to trigger the avalanche.

By the definition of capacitance, $Q_{S}$ can be expressed as \cref{eq:6}.
\begin{equation}
    \label{eq:6}
    Q_{S} = C_{S} \cdot (V_{bias} - V_{bd})
\end{equation} where $C_{S}$ symbolizes the equivalent capacitance of SPAD, $V_{bias}$ means the applied reversed voltage, and $V_{bd}$ indicates the breakdown voltage of SPAD.

Through reintroducing \cref{eq:6} into \cref{eq:5}, the gain variation of a GTM readout channel ($\Delta G_{c}$) can be expressed as \cref{eq:7}, which is proportional to the change in applied voltage.
\begin{equation}
    \label{eq:7}
    \Delta G_{c} \propto {\rm \sum_{fired}}^{} \Delta G_{S} \propto \Delta V_{bias}
\end{equation} where ${\rm \sum_{fired}}^{}$ states the sum over all fired SPADs within SiPMs of a channel.

\subsection{Temperature Dependence}
\label{sec:5.3}

Similar to \cref{sec:5.2}, \cref{fig:13.2} displays the gain as a function of temperature (T). The distinction lies in the fact that the gain variation is inversely proportional to the change in temperature on the circuit board. At a fixed voltage, $Q_{S}$ in \cref{eq:6} is influenced by $V_{bd}$, which is actually a function of temperature \citep{Hofbauer_2018}. Hence, \cref{eq:7} can be rewritten as \cref{eq:8}.
\begin{equation}
    \label{eq:8}
    \Delta G_{c} \propto - \Delta V_{bd\ (T)}
\end{equation} 

The main reason for the change in $V_{bd}$ with temperature is the alteration in the band gap energy ($E_{g}$). As the temperature increases, $E_{g}$ decreases \citep{Varshni_1967}, subsequently reducing the minimum energy required to ionize secondary electrons and holes in the depletion region \citep{Anderson_1972}. In essence, this implies an increase in the number of collisions of the first excited electron in the depletion region. Consequently, a higher $V_{bd}$ is needed to supply enough kinetic energy to trigger the avalanche effect.

Generally speaking, the change in $E_{g}$ is inversely proportional to the square of the temperature change at lower temperatures and inversely proportional to the temperature change at higher temperatures. The temperature at which this phase transition occurs varies depending on the material. Another study has shown that $V_{bd}$ of SiPM increases linearly with temperature over the operating temperature range \citep{Otte_2017}, so the gain of each channel can be well-fitted by a line with a negative slope, as demonstrated in \cref{eq:9}.
\begin{equation}
    \label{eq:9}
    \Delta G_{c} \propto - \Delta T
\end{equation}

\section{Conclusion}
\label{sec:6}

In this study, the GRB saturation fluence of GTM was estimated through simulation and detailed knowledge of the instrument. During energy calibration using various radioactive isotopes and \ce{LYSO(Ce)}, a crosstalk issue was identified in the gamma-ray telescope made with pixelated scintillators coupled with multi-channel SiPMs. Additionally, the SiPM's saturation effect was observed and resolved when measuring isotopes with peaks higher than 1 MeV. After discarding the relatively weak channel noise and applying proper ADC-to-energy mapping, all measurement data spectra were successfully recovered. Using the corrected spectra, the energy resolution and variations in gain with voltage and temperature of GTM were determined. These results will be very helpful for the future operation of the GTM mission. By knowing the operating temperature of GTM in space, the appropriate applied voltage can be regulated based on the predicted temperature gain. With both temperature and voltage gains known, the measured ADC can be converted back to energy for subsequent spectral analysis, which the energy resolution results can provide a rough understanding of the error.

\section*{Acknowledgment}
This work is supported by Taiwan Space Agency (TASA) under grant NSPO-P-109221 and the National Science and Technology Council (NSTC) of Taiwan under grant 112-2112-007-053. C.-P.H. acknowledges support from the NSTC in Taiwan through grant 112-2112-M-018-004-MY3.

\bibliographystyle{elsarticle-num}
\bibliography{references}

\end{document}